# Gravitationally redshifted absorption lines in the X-ray burst spectra of a neutron star

J. Cottam*, F. Paerels†, & M. Mendez‡

*NASA Goddard Space Flight Center, Laboratory for High Energy Astrophysics, Greenbelt, MD, 20771, USA

†Columbia Astrophysics Laboratory and Department of Astronomy, Columbia University, 538 W. 120th St., New York, NY 10027, USA

‡SRON National Institute for Space Research, Sorbonnelaan 2, 3584 CA Utrecht, the Netherlands

**The fundamental properties of neutron stars provide a direct test of the equation of state of cold nuclear matter, a relationship between pressure and density that is determined by the physics of the strong interactions between the particles that constitute the star. The most straightforward method of determining these properties is by measuring the gravitational redshift of spectral lines produced in the neutron star photosphere[1]. The equation of state implies a mass-radius relation, while a measurement of the gravitational redshift at the surface of a neutron star provides a direct constraint on the mass-to-radius ratio. Here we report the discovery of significant absorption lines in the spectra of 28 bursts of the low-mass X-ray binary EXO 0748−676. We identify the most significant features with the Fe XXVI and XXV $n = 2−3$ and O VIII $n = 1−2$ transitions, all with a redshift of $z = 0.35$, identical within small uncertainties for the respective transitions. For an astrophysically plausible range of masses ($M \sim 1.3−2.0 \, M_\odot$)[2-5], this value is completely consistent with models of neutron stars composed of normal nuclear matter, while it excludes some models[6-7] in which the neutron stars are made of more exotic matter.**

The XMM-*Newton* observatory[8] observed the Low Mass X-ray Binary EXO 0748−676[9] during its Commissioning and Calibration phases for almost a half million seconds, spread over six satellite orbits between 2000 February 21 and 2000 April 21. Data were recorded with the Reflection Grating Spectrometer[10](RGS) for 335,000 seconds (data obtained with the European Photon Imaging Cameras[11,12] (EPIC) are available for 39,000 seconds of simultaneous exposure). During this time, a total of 28 X-ray bursts were recorded with the RGS, lasting a cumulative 3200 seconds. During the brief bursts, the neutron star outshines the accretion-generated light by an order of magnitude in intensity, while the ongoing accretion ensures a continuing supply of heavy elements in the stellar photosphere. This makes the burst spectrum a promising place to detect absorption structure from a neutron star photosphere, a long standing goal in compact object astrophysics. With a detailed stellar photospheric spectrum, the techniques of classical stellar spectroscopy could be employed to measure the fundamental parameters of neutron stars. It should also be possible to measure the general relativistic gravitational redshift, which provides additional constraints on the mass and radius of the star. Due to the long exposure, and the high efficiency and spectral resolving power of the RGS, this EXO 0748−676 dataset is by far the most sensitive to date for conducting such a search.

Data were processed with the XMM-*Newton* Science Analysis Software (SAS) that is currently available as version 5.3.3. The soft X-ray lightcurve of EXO 0748−676 shows considerable variability[13]. Searching for X-ray bursts in the RGS lightcurve, we only considered events which conformed to the steep rise/exponential decay shape characteristic of type I X-ray bursts[14]. We identified 28 bursts in the RGS data. The bursts varied in peak intensity between 4 cts s$^{-1}$ and 12 cts s$^{-1}$ with an average of 8.8 cts s$^{-1}$. This represents an increase by a factor of ~15 over the quiescent levels observed in the periods of low activity. We defined the onset of a burst as the time at which the count rate first rises above the quiescent level by a factor of two or more. The end, less well defined, occurs when the count rate drops back to the local average level. The majority of the bursts ranged in duration from 48 to 128 seconds, with an average of 90 seconds. Seven of the bursts were longer, with durations between 176 and 320 seconds.

We then extracted the first order ($m$ = -1) RGS spectra for each burst. The spacecraft pointing was stable during observations on each of the separate revolutions, but differed between revolutions by up to 40 arcsec. To generate the average burst spectrum we therefore combined the data for all observations within a single revolution, but generated separate spectral files and response matrices for each revolution. All spectral fitting was performed simultaneously on these separate data sets. For ease of display we generated a flux spectrum of the average burst, using the effective area curves for each separate data set. This allowed us to combine the data from the two RGS instruments as well. The wavelength scale is accurate to ~10 mÅ. The effective area is accurate to 5% for all wavelengths longer than 8 Å.[10] Background subtraction was performed using the same extraction algorithms, but over the image region not occupied by the source. The background flux becomes a significant fraction of the total flux for wavelengths longer than ~32 Å. We therefore only consider the wavelength range from 8 to 32 Å in our analysis.

To constrain the broad-band properties of the burst spectrum we examined the EPIC data. Pile-up during the bright bursts contaminated all but 250 seconds, or three bursts. The EPIC/PN spectrum of these three bursts is well fit by a blackbody, with a peak color temperature of $kT_{BB}$ ~ 1.8 keV, decaying to $kT_{BB} \leq 1.5$ keV. Since the spectral properties clearly evolve during the bursts, we investigated the RGS spectrum of the early, bright phases and the decay phases separately. We explored a variety of ways to subdivide the bursts, but found that the results are not sensitive to the exact criterion used to compile the 'early' and 'late' time burst spectrum. We therefore chose to divide the bursts in the simplest way, by splitting them in half by duration. The resulting flux spectra for the early and late phases of the averaged burst are shown in Figure 1.

As in the case of the quiescent spectrum,[13] we see clear evidence for absorption and emission from highly ionized gas surrounding the neutron star during the bursts. O VII K-shell emission (consisting of $n = 1-2$ resonance, intercombination, and forbidden lines at 21.60, 21.80, and 22.10 Å) is clearly detected, as is absorption by O VII (photoelectric absorption at 16.78 Å, resonance line absorption at 21.60 Å). The fact that the O VII line emission is dominated by the intercombination transition indicates that the gas is recombining, which implies that the ionization is driven by photoionization, and that the electron density is relatively high ($n \geq 10^{12}$ cm$^{-3}$). We see a clear change in the nature of the O VII spectrum as the bursts progress, with the emission weakening and the absorption increasing from the early to the late phases, indicative of an overall progressive flattening of the source geometry towards the line of sight. Wavelength shifts in the O VII absorption features indicate a significant bulk outflow velocity of $v \sim 5000$ km s$^{-1}$ during the bursts.

In order to develop a physically consistent model for the spectral transmission of the circumstellar absorber, which we need in order to quantitatively account for its contribution to the observed absorption spectra, we fit the O VII and O VIII spectra and measured the intensity, velocity width and Doppler-shifts of the emission lines, and the ion column density and velocity broadening of the absorption features for both the early and late phase spectra. The measurements assume an empirical continuum spectrum during the bursts, constructed from a blackbody and a powerlaw, whose parameters were optimized by eye. O VII and O VIII absorption spectra were calculated with a spectral code originally developed to interpret the X-ray absorption spectra of AGN.[15] The model incorporates atomic structure and transition probabilities, and for any given ion, consistently accounts for the absorption in all transitions out to high principal quantum number and the photoelectric continuum.  The entire spectrum was subject to absorption by a neutral medium, with equivalent hydrogen column density $N_H = 1 \times 10^{21}$ cm$^{-2}$; the neutral absorption spectrum has strong O K absorption in the 22−24 Å range, the shape of which we optimized to conform to the interstellar O absorption spectrum measured in other sources.[16]

We then synthesized a model for the full spectral transmission of the circumstellar absorber.  We adopted an ionization parameter, $\xi \equiv L_{ionizing}/nR^2$ (with $L_{ionizing}$ the ionizing luminosity, $n$ the density of the medium, $R$ its distance to the ionizing source[17,18]), of $\xi = 10$ as representative, so as not to overproduce O VIII absorption, and added the other elements at their solar abundances, with ionization fractions derived from the photoionization equilibrium balance. The full set of ions includes the K-shell ions of C, N, O, Ne, Mg, and Si, and the L-shell ions of Fe. We scaled the turbulent velocity broadening of each ion with the value measured in the O VII resonance line, assuming a common temperature for all ions.  We adopted the Doppler blueshift observed in the O VII features for all ions. The resulting transmission model, superimposed on the optimized continuum model, is overplotted on the observed spectra in Figure 1. The apparent absence of λ 24.78 Å N VII Lyα absorption in the data implies a subsolar N/O ratio in the absorbing gas. The absence in the data of absorption by ions that are present at higher ionization parameter – specifically the absence of significant Fe L absorption – implies that the circumstellar medium occupies only a narrow range of (fairly low) ionization parameters.

We now examine the spectrum for any remaining structure that is not associated with the circumstellar absorber.  In the early time burst spectrum, the most significant modulation appears at 13.0 Å. We also see weaker structure at 25.3, 26.3, and 26.9 Å. In the late time spectrum, we identify significant modulations at 13.75, 25.2, and 26.4 Å with weak features at 17.8 and 19.7 Å.  In view of the noise levels, it is difficult to perform such a search effectively using statistical significance criteria only. We will therefore appeal to spectroscopic consistency arguments when assessing apparent absorption features in the spectrum. As a guide, however, we estimated the significance of the features by fitting each with a simple Gaussian profile, optimizing the continuum, and evaluating their significance in terms of the amplitude of the Gaussian. In the early phase spectrum, the depth of the 13.0 Å feature differs by 7 σ from zero, while those of the 25.3, 26.3, and 26.9 Å features differ by 3-4 σ. In the late phase spectrum, the 13.75 Å feature has a 5 σ amplitude, and the 25.2 and 26.4 Å features each have amplitudes of 3 σ. The features at 17.8 and 19.7 Å in the late phase spectrum have amplitudes less than 1 σ, and we therefore exclude these features from further consideration.

We inspected the spectra obtained by the two RGS spectrometers separately and find that all the features appear in both parallel spectra, but with only marginal detections in the case of the 25.3, 26.3, and 26.9 Å (early times) features. The fact that all significant features appear either at early or at late times rules out the possibility of stationary modulations in the spectrometer efficiency. We have also examined the three very deep RGS spectra of extragalactic featureless continuum sources (Mkn 421, 3C273, and PKS2155−304; Rasmussen et al., in preparation). No significant modulations are observed in these spectra at the positions of the features under discussion, nor do these spectra exhibit unexplained features at any other wavelength, of strength comparable to the ones observed here.

Neither do the features match the wavelengths of absorption lines expected from the circumstellar absorber, with the possible exception of the 26.4 Å feature in the late phase spectrum, which is probably partly due to CVI Lyγ at 26.6 Å. We examined absorption models appropriate to higher ionization parameters (over the range $\xi =$ 10−100) and found that the features cannot be made consistent with circumstellar absorption at any ionization parameter. This explanation can be rejected on purely spectroscopic grounds: it requires significant velocity shifts that vary randomly between ions that are present at similar ionization parameters. Furthermore, fitting individual absorption lines to the features produced serious inconsistencies in terms of the predicted overall absorption structure due to any given single ion. Finally, the features do not appear in the (accretion-dominated) quiescent spectrum of the source.

We are left with the exciting possibility that the absorption features arise in the photosphere of the neutron star. The magnetic fields in the neutron stars in low mass X-ray binaries are believed to be small[19-21], so that field effects do not affect the atomic structure, and we can use well established atomic spectroscopy to interpret the wavelengths. In the only other example of nontrivial structure in the spectrum of a neutron star,[22] the correct spectroscopic identification, and hence the inferred redshift, depends critically on the unknown strength of the stellar magnetic field. We note that the previous detections of large equivalent width features in the 4-5 keV band with proportional counters[23] have not been substantiated by subsequent observations,[24] and these and other similar detections,[25-26] are most likely due to instrumental effects.

We have examined all the spectra of the K-shell ions of C through Si, and find no multiple coincidences between observed and predicted line positions, with identical redshifts for all ions. However, an important clue to identifying the features lies in the presence of the 13.0 Å line at early times and the presence of the 13.75 Å line at late times. A solution to the Saha ionization balance for densities $n \sim 10^{23}$ cm$^{-3}$ (as expected in a neutron star atmosphere at Rosseland optical depth unity, above which most of the absorption spectrum is formed) indicates that iron should be primarily in its H-like charge state at temperatures $kT \geq 1.2$ keV, while at $kT \leq 1.2$ keV, the He-like charge state dominates; these temperatures roughly correspond to the observed color temperature early and late in the bursts, respectively. At these temperatures, the ionization balance of iron does not shift into the fully stripped charge state until the density drops below $\sim 10^{22}$ cm$^{-3}$. The density and temperature in a real atmosphere exhibit significant gradients, but the Saha balance at a representative temperature and density usually provides a good indication as to which ions of a given element are likely to be dominant in the stellar absorption spectrum. The $n = 2-3$ transitions in the H-like ion (the analogue of the H$\alpha$ Balmer line), occur at 9.518, 9.533, 9.579 and 9.675, 9.690, 9.738 Å.[27] Identifying these transitions with the feature at 13.0 Å in the early phase burst spectrum implies a redshift of $z = 0.35$. The strongest $n = 2-3$ transitions in the He-like ion occur at 10.213, 10.048 Å (Ehud Behar, priv. comm.; the 10.213 Å transition has the highest oscillator strength). Identifying these transitions with the feature at 13.75 Å in the late phase spectrum also implies a redshift of $z = 0.35$. The higher order series members will all lie at wavelengths shortward of 10 Å, to which our spectrum is not sensitive.

The Saha balance at temperatures $kT \geq 1$ keV and densities $n \sim 10^{23}$ cm$^{-3}$ indicates that all the lighter elements should be nearly stripped, so we would not expect to see their K-shell absorption lines, with the possible exception of oxygen, in view of its relatively high abundance. Applying the same redshift to the O VIII Ly$\alpha$ line (rest wavelength 18.97 Å) we would expect to see a feature at 25.6 Å. We speculate that the double $\lambda\lambda$ 25.2, 26.4 Å structure observed at late phases, which is centered at this wavelength, is in fact a self-reversed, broad O VIII Ly$\alpha$ line, where the self-reversed profile is indicative of extended structure to the outer atmosphere, and possibly a slow outflow, such as observed in the strong UV resonance lines in massive stars with extended atmospheres.[28] The corresponding O VIII Ly$\beta$ line (21.60 Å at $z = 0.35$) would be hidden in the O VII spectrum from the circumstellar medium. The remaining features are all of relatively low significance, have no obvious spectroscopic interpretation, and are most likely statistical fluctuations. A quantitative interpretation of the strength and shape of the features we have identified will have to await a dedicated full model atmosphere calculation, which may have to incorporate effects induced by the X-ray bursts.

We have identified three sets of redshifted transitions in iron and oxygen in the EXO 0748-676 spectrum, all with an implied redshift of $z = 0.35$. We have compared our result with the family of theoretical mass-radius relations collected by Lattimer and Prakash.[6] A redshift of $z = 0.35$ is consistent with most modern equations of state for neutron stars composed of normal matter in the mass range of $M \sim 1.4-1.8 \, M_\odot$ and the radius range of $R \sim 9$-12 km, depending on the choice of $M$-$R$ relation. This is compatible with a neutron star with a birth mass near the average for non-accreting neutron stars ($M = 1.4 \, M_\odot$), that has been accreting mass at the observed rate for $\sim 10^9$ years, and agrees with the estimated masses of other accreting neutron stars.[2-4] A redshift of $z = 0.35$ is inconsistent with several mass-radius relations based on equations of state for more exotic matter such as strange quark matter or kaon condensates,[6-7] unless, for some of these equations of state, the mass of the neutron star in EXO 0748-676 is $\leq 1.1 \, M_\odot$, which is uncomfortably low from astrophysical arguments.

Acknowledgements

This work is based on observations obtained with the XMM-*Newton*, an ESA science mission with instruments and contributions directly funded by ESA member states and the USA (NASA).  We thank Ehud Behar for supplying us with results from his atomic structure calculations of the He-like Fe ion, and Masao Sako for the use of his absorption spectral code.

Correspondence and requests for materials should be addressed to J.C. (e-mail: jcottam@milkyway.gsfc.nasa.gov).

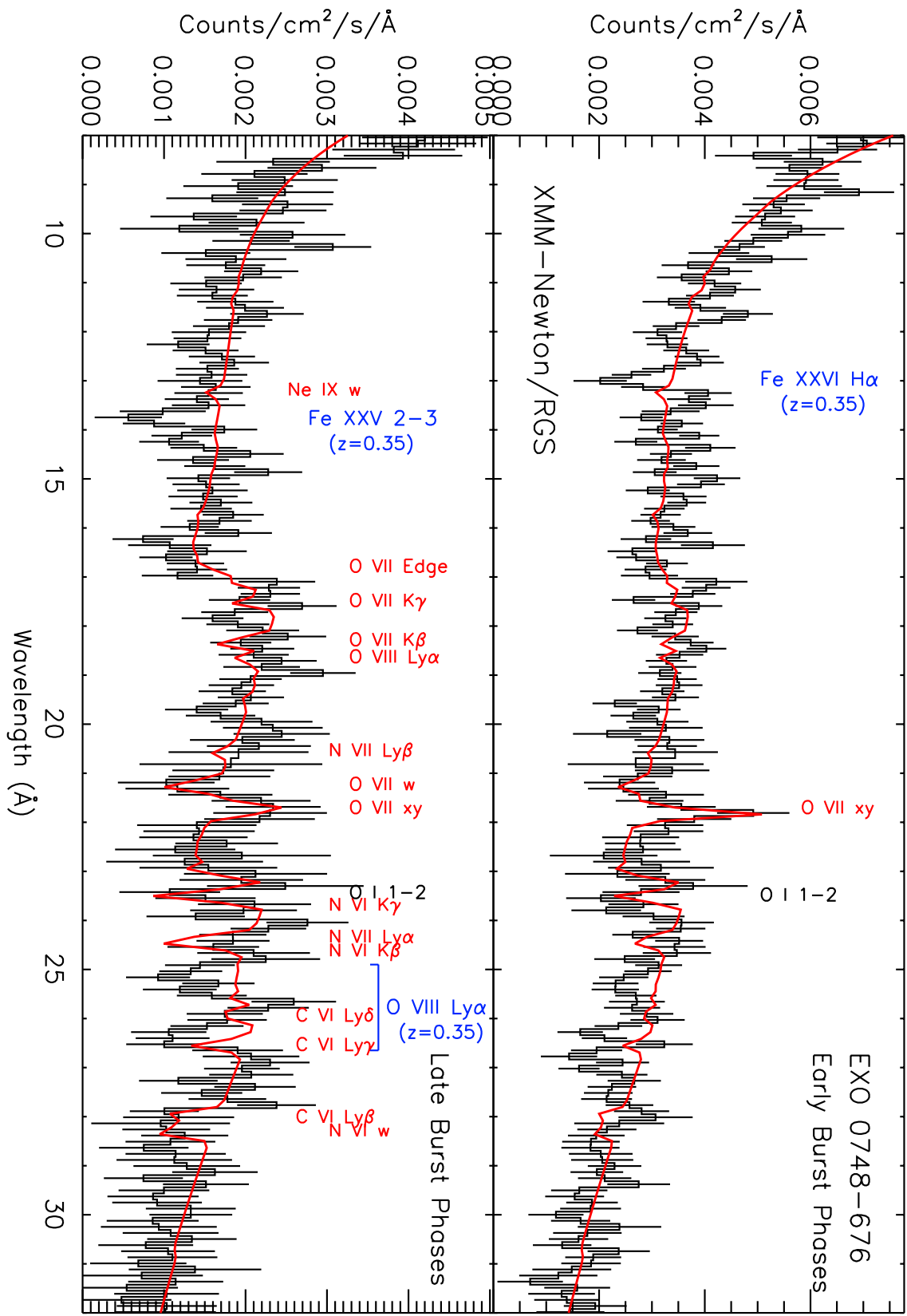

Figure 1: The XMM-Newton RGS spectra of EXO 0748-676 for 28 type I X-ray bursts. The background-subtracted flux spectra for the early and late phases of the bursts are shown in the top and bottom panels respectively. The data are plotted as the black histograms, with $1\sigma$ error bars derived from counting statistics. The red line is the empirical continuum, with additional O VII intercombination line emission, modulated by absorption in photoionized circumstellar material. In red, we have labeled the positions of the most prominent discrete absorption lines from the circumstellar medium; in the He-like spectra, 'w' signifies the $n$ = 1−2 resonance transition, 'xy' the (unresolved) $n$ = 1−2 intercombination transitions, while higher series members are marked 'K$\beta,\gamma$', etc. Column densities in ions other than O VII have been normalized to the absorption measured in O VII, assuming ionization parameter $\xi$ = 10, and solar abundances. The N VII Ly$\alpha$ line at 24.78 Å is overpredicted, indicating a subsolar N/O abundance ratio. The black labels indicate the interstellar O 1s−2p absorption line. Blue labels indicate the photospheric absorption lines in Fe XXVI, XXV, and O VIII, at a redshift $z$ = 0.35. The data and models have been rebinned to $\Delta\lambda$ = 0.124 Å, which is a factor of ~ 2.5 larger than the RGS instrument resolution.